# The GenoChip: A New Tool for Genetic Anthropology

Eran Elhaik[1], Elliott Greenspan[2], Sean Staats[2], Thomas Krahn[2], Chris Tyler-Smith[3], Yali Xue[3], Sergio Tofanelli[4], Paolo Francalacci[5], Francesco Cucca[6], Luca Pagani[3,7], Li Jin[8], Hui Li[8], Theodore G. Schurr[9], Bennett Greenspan[2], R. Spencer Wells[10,*] and the Genographic Consortium

[1] Department of Mental Health, Johns Hopkins University Bloomberg School of Public Health, 615 N. Wolfe Street, Baltimore, MD 21205, USA
[2] Family Tree DNA, Houston, TX 77008, USA
[3] The Wellcome Trust Sanger Institute, Wellcome Trust Genome Campus, Hinxton, UK
[4] Department of Biology, University of Pisa, Italy
[5] Department of Natural and Environmental Science, Evolutionary Genetics Lab, University of Sassari, Italy
[6] National Research Council, Monserrato, Italy
[7] Division of Biological Anthropology, University of Cambridge, UK
[8] Fudan University, Shanghai, China
[9] University of Pennsylvania, Philadelphia, PA
[10] National Geographic Society, Washington DC, USA

*Please address all correspondence to Spencer Wells at spwells@ngs.org

**Abstract**

The Genographic Project is an international effort using genetic data to chart human migratory history. The project is non-profit and non-medical, and through its Legacy Fund supports locally led efforts to preserve indigenous and traditional cultures. While the first phase of the project was focused primarily on uniparentally-inherited markers on the Y-chromosome and mitochondrial DNA, the next is focusing on markers from across the entire genome to obtain a more complete understanding of human genetic variation. In this regard, genomic admixture is one of the most crucial tools that will help us to analyze the genetic makeup and shared history of human populations. Although many commercial arrays exist for genome-wide SNP genotyping, they were designed for medical genetic studies and contain medically related markers that are not appropriate for global population genetic studies. GenoChip, the Genographic Project's new genotyping array, was designed to resolve these issues and enable higher-resolution research into outstanding questions in genetic anthropology. We developed novel methods to identify ancestry informative markers (AIMs) and genomic regions that may be enriched with alleles shared with ancestral hominins. Overall, we collected and ascertained AIMs from over 450 populations. Containing an unprecedented number of Y-chromosomal and mtDNA SNPs and over 130,000 SNPs from the autosomes and X-chromosome, the chip was carefully vetted to avoid inclusion of medically relevant markers. The GenoChip results were successfully validated by genotyping more than 500 samples from the 1000 Genomes Project and private collections. To demonstrate its capabilities, we compared the $F_{ST}$ distributions of GenoChip SNPs to those of two commercial arrays for three continental populations. While all arrays yielded similarly shaped (inverse J) $F_{ST}$ distributions, the GenoChip autosomal and X-chromosomal distributions had the highest mean $F_{ST}$ (0.10 and 0.13, respectively), attesting to its ability to discern subpopulations. In summary, the GenoChip is a dedicated genotyping platform for genetic anthropology and promises to be the most powerful tool available for assessing population structure and migration history.

## Introduction

Apportionment of human genetic variation has long established that all living humans are related via recent common ancestors who lived in sub-Saharan Africa about 200,000 years ago [1]. The world outside Africa was settled over the past 50,000-100,000 years [2,3] when the descendents of our African forebearers spread out to populate other continents [4]. While genome-wide data depict a single major dispersal of modern humans from Africa, recent findings suggest at least two archaic admixture (interbreeding) events with extinct hominins, one with Neanderthals in Eurasia and the second with Denisovans in Southeast Asia [5,6]. The recurrent interbreeding, admixture, and migration events shaped modern populations into mosaics of ancient and recent haplotypes – varying only by the proportion of their ancestral genetic blocks, not in the building blocks themselves. Passed down the sex lines nearly unaltered from one generation to the next, these blocks carry the story of our most recent common ancestors' journeys. Although highly similar, the subtle genetic differences between populations are like breadcrumbs allowing us to trace human past through mixing events and migrations.

The advent of next-generation sequencing technology that enabled the rapid genotyping of hundreds of thousands of genetic markers revolutionized genetic anthropology and launched several human sequencing projects, in the process broadening our understanding of genetic diversity and migration history [e.g., 7,8,9,10]. The rapidly escalating interest in ancestral histories of foreign continents and geographical regions, and the greater antiquity of populations from Africa [11], India [12], the Americas [13], the Caucasus [14], and Oceania [15], have already yielded deeper splits among these regions and complex patterns of past migration and admixture events.

Despite this tremendous progress, many unresolved problems hinder progress in the field. First, only a handful of the estimated 6,000 global population groups were genotyped and studied, which limits the phylogeographic resolution of the findings. Second, the plethora of genetic markers obtained from different genotyping platforms has resurrected the “empty matrix” problem, whereby populations from different studies can barely be compared. Finally, genotyping costs remained prohibitively high and unjustified as commercial genotyping platforms do not accommodate crucial ancestry informative markers (AIMs) [16]. Furthermore, these arrays are also enriched in trait- or disease-related markers, which prompts a host of psychological, social, legal, political, and ethical concerns from the individual, population, and global levels [17].

The first phase of the Genographic Project focused on reconstructing human migration through the analysis of uniparentally-inherited markers on the Y-chromosome and mitochondrial DNA. This phase yielded three dozen publications that described the population structure of modern populations [e.g., 14], traced migration routes of ancient populations [e.g., 18], improved the resolution of Y and mtDNA phylogenetic trees [e.g., 19,20,21], and attempted to identify genetic relatedness to historical figures [e.g., 22,23]. The project has not only been successful in inferring details of human migration histories, but also in attracting public participants interested in testing their own DNA to better understand their genetic ancestry. As of December 2012, over 500,000 public participants from more than 130 countries have purchased Genographic public participation kits, of which over 200,000 have submitted their results to an anonymous database to be used for research purposes.

The success of the Genographic Project has also helped to spawn a broader interest in what has been dubbed "genetic genealogy." Unlike other fields of science where the knowledge and excitement of new discoveries are often confined to members of the scientific community, genealogy, and particularly genetic genealogy, has a very large, devoted, supportive, and highly active public community. Since the 1970s, genealogy has become one of the most popular avocations, enjoyed by millions of people around the world seeking to trace their ancestry and learn about the geographical regions in which their ancestors lived [24]. While traditional genealogy requires written records, which only extend several generations into the past at most, genetic genealogy is easily available to everyone in the form of direct-to-consumer genetic ancestry tests and friendly computational interfaces that visualize and explain the results [25]. The increasing fascination with genetics has prompted entrepreneurs to offer self-test kits and provide information ranging from disease risk and life-style choices to genetic ancestry [26]. Some of the existing solutions have been criticized for making deceptive health-related claims and for providing limited and imprecise answers regarding ancestry [17]. The concerns about ancestry reporting were not unjustified, as these entrepreneurs adopted problematic commercial genotyping platforms for assessing genetic background.

While the uniparental markers used in the first phase of the Genographic Project remain useful for studies of human migration history, many questions cannot be answered without assessing patterns of genetic variation in the rest of the genome. Legal and ethical obstacles often hampered autosomal-driven studies due to the difficulty in discerning markers that are informative for genetic anthropology from those having medical relevance. Since the launch of the Genographic Project and similar endeavors, a growing number of individuals and populations have expressed serious concerns about the sharing and potential exploitation of their medical information [17,27]. Unfortunately, all commercially available genotyping platforms were optimized for medical genetic work, making them unsuitable for population genetic studies due to privacy concerns and biased allele frequencies. Even the exceptional Affymetrix Human Origins array, which was also designed for population genetic studies [28], not only contains tens of thousands of coding and medically-related markers, but was ascertained based on data from only 12 populations, which may limit its applicability to study worldwide genetic diversity. To reconcile the competing interests in investigating population genetic history and identifying variants contributing to health, we decided to develop a novel SNP array – which we call the GenoChip.

Our goals were to design, manufacture, validate, and test a state of the art SNP array dedicated solely to genetic anthropology and to genotype a large number of samples and populations to learn more about human history. The GenoChip is an Illumina HD iSelect genotyping bead array designed with over 130,000 highly informative autosomal and X-chromosomal markers, ~12,000 Y-chromosomal markers, and ~3,300 mtDNA markers. As AIMs are invaluable tools in population genetics and genetic anthropology for discerning subpopulations, we focused our efforts in collecting the most informative ones [29]. The autosomal and X-chromosomal marker sets comprised of over 75,000 AIMs were ascertained from over 450 worldwide populations, making it the largest and most detailed AIMs panel ever constructed (Figure 1). Half of our AIMs were culled from the literature and private data sets and the rest were calculated using *infocalc* [30] and *AIMsFinder*. We applied these two methods on global panels comprised of nearly 300 populations assembled from public and private data sets. Many of these populations are unique to our project and have never before studied or searched for AIMs. We used *infocalc* to identify AIMs in population panels organized by the source of the data, whereas *AIMsFinder* was applied in a pairwise fashion over all populations. We excluded SNPs in high LD ($r^2$>0.4) in all populations, except for hunter-gatherer like the Hadza and Sandawe of Tanzania [31] and Melanesian populations [10]. To determine the extent of gene flow from Neanderthal and Denisovan to modern humans, we collected from the literature [5,32,33] SNPs and haplotypes from genomic regions bearing evidence of interbreeding. In addition, we used a modified version of *IsoPlotter* [34] to identify regions in which modern humans and Neanderthals share the derived allele and chimpanzees and Denisovans share the ancestral allele. Using the same approach, we identified SNPs within regions enriched for the Denisovan sharing derived alleles with modern humans. Overall, nearly 30,000 SNPs were collected in these interbreeding hotspots. We also included ~1,400 high-confidence Paleo-Eskimo Saqqaq SNPs [35] and over 13,000 high-confidence Aboriginal Australian SNPs [15]. Finally, we randomly selected 3,000 common HapMap III SNPs (minor allele frequency [MAF]>20%) and 4,500 1000 Genomes Project SNPs with an allele frequency of at least 10% with at least one continental population. To prevent false-positives, we included mostly SNPs observed in the HapMap III and the 1000 Genome Project data sets [7,9]. We further eliminated A/T and C/G SNPs to minimize strand misidentification.

Overall, the GenoChip is highly compatible with other commercial arrays. Some 76% of our SNPs overlap with those in the Illumina Human 660W-Quad array, 55% overlap with the Illumina HumanOmni1-Quad, Illumina Express, and Affymetrix 6.0 arrays, and 40% overlap with the Affymetrix 5.0 and Affymetrix Human Origins arrays. In addition, we constructed over 45,000 probes to identify SNPs defining all known Y-chromosome and mtDNA haplogroups, many of which were not reported in the literature (Supplementary Texts 1-2). The resulting chip has a SNP density of at least one per hundred kilobases over 92% of the assembled human genome (hg19) (Figure 2), including regions uncharted by the HapMap (I-III) and HGDP projects [9,10,36]. We further note that this choice of SNPs is particularly suitable for imputation. We thus designed an array suitable to detect gene flow from ancient hominins into modern humans as well as between the putatively oldest populations, such as hunter-gatherers (e.g.,Khoi-San) and more recent populations, such as the Druze or European Jews [37]. In this regard, the GenoChip differs from alternative solutions geared toward analyzing far fewer populations and uniparental markers that may lack geographic specificity.

Several steps were taken to ensure that the genetic results would not be exploited for pharmaceutical, medical, and biotechnology purposes. First, participant samples were maintained in a completely anonymous status during GenoChip analysis. Second, no phenotypic or medical data were collected from the participants. Third, we included only SNPs in noncoding regions without any known functional association, as reported in dbSNP build 132. Lastly, we filtered our SNP collection against a 1.5 million SNP data set containing all variants that have potential, known, or suspected associations with diseases.

To construct this dataset, we extracted SNPs from multiple open-access databases including OMIM (http://www.ncbi.nlm.nih.gov/omim/), the Cancer Genome Atlas [38], PhenCode [39], NHGRI GWAS Catalog [40], The Genetic Association Database [41], MutaGeneSys [42], GWAS Central [43], and SNPedia [44], as well as SNPs identified in the MHC region. We also excluded SNPs reported to be associated with Phenotypic traits. Finally, to circumvent imputation efforts toward inferring potential medical-relevant SNPs, we excluded SNPs that were in high LD ($r^2$>0.8) with any of the other SNPs included in our dataset. Overall, we assembled nearly 170,000 SNPs dedicated for genetic anthropological and genealogical research without any known health, medical, or phenotypic relevance (Table S1).

We validated the results of the GenoChip by genotyping 168 worldwide samples from the 1000 Genomes Project and cross-validating the autosomal genotypes. The concordance rate per sample was over 99.5%. The marginal error rate was expected due to the low coverage of the 1000 Genomes Project data, particularly for rare alleles [7]. An additional 400 samples were genotyped to test the ability of the GenoChip to infer Y-chromosome and mtDNA haplogroups. The average success rates for the paternal and maternal haplogroups were 82% and 90%, respectively (Figure 3). It is very likely that the GenoChip can capture many of the remaining haplogroups that are missing from our reference set, and, for this reason, we are currently conducting a larger genotyping effort to validate them. Overall, we confirmed that GenoChip produces highly accurate results and has broad coverage for SNPs defining Y-chromosome and mtDNA haplogroups.

To demonstrate that the GenoChip consists of highly informative SNPs suitable for population genetic studies, we compared its performance to that of the Illumina Human660W and Affymetrix Human Origins arrays. A comparison of the minor allele frequency (MAF) distributions between the three arrays revealed gross differences in allele frequencies (Figure 4, S1). Due to the high frequency of rare alleles in the HapMap dataset, none of the arrays resembled the shape of the HapMap's MAF distribution. Nonetheless, the commercial arrays were enriched in rare autosomal SNPs compared to the GenoChip. Similar findings were observed for X-chromosome SNPs, although to a lesser extent, as the GenoChip consists of a higher fraction of extremely rare SNPs as well as common SNPs. The different MAF distributions correspond to the choices of SNP ascertainment made in each project.

To assess the extent of genetic diversity that can be inferred among human subpopulation by the different arrays, we next compared their $F_{ST}$ distributions [45,46,47]. $F_{ST}$, measures the differentiation of a subpopulation relative to the total population, and is directly related to the variance in allele frequency between subpopulations, such that a high $F_{ST}$ corresponds to a larger difference between subpopulations [48]. Elhaik [49] used 1 million markers that were genotyped in eight HapMap populations (YRI, LWK, MKK, CEU, TSI, CHB, CHD, and JPT) to carry out a two-level hierarchical $F_{ST}$ analysis. He showed that the greatest proportion of genetic variation occurred within individuals residing in the same populations, with only a small amount (12%) of the total genetic variation being distributed between continental populations and even a lesser amount (1%) between intra-continental populations. An $F_{ST}$ distribution for three continental populations employing 3 million HapMap SNPs yielded an even lower estimate (0.08) to the proportion of genetic variation distributed between continental populations due to the large number of rare alleles [49].

$F_{ST}$ distributions for three continental populations were similarly calculated for GenoChip and the two commercial arrays. Although all $F_{ST}$ distributions were similar in shape to the HapMap $F_{ST}$ distribution, they differed in their means (Figure 5, S2). For autosomes and X-chromosomal SNPs, the mean $F_{ST}$ of both Illumina Human660W and Human Originals were lower than that of the GenoChip due to the high fraction of rare uninformative SNPs in the two commercial arrays. These results may suggest a reduced ability to infer common SNPs that are likely old and correspond to ancient demographic and natural processes [50,51].

The Illumina Human 660W array had the highest fraction of low-$F_{ST}$ alleles, suggesting it is less suitable for population genetic studies compared to the GenoChip and Human Origins. As only half of the Human Origins SNPs could be tested, it is difficult to evaluate its performance. However, we speculate that the large number of rare SNPs, along with those not studied, which are also likely to be rare, may reflect the small number of populations used for its ascertainment and the number of alleles private to these populations. Because the MAF and $F_{ST}$ were not used as filtering criteria for the GenoChip SNPs, we can conclude that its enrichment toward high-$F_{ST}$

SNPs mirrors the success of our ascertainment process, emphasizing its potential for population genetic studies.

To summarize, we designed, developed, validated, and tested the GenoChip, the first genotyping chip completely dedicated to genetic anthropology. The GenoChip will help to clarify the genetic relationships between archaic hominins such as Neanderthal and Denisovan, and modern humans, and provide a more detailed understanding of human migration history. We compared the GenoChip to two commercially available arrays and demonstrated the superior ability of the GenoChip to differentiate subpopulations within global data sets. For these reasons, we expect the expanded use of the GenoChip in genetic anthropology research will yield important new insights into the history of our species.

## Acknowledgments

We are grateful to David Reich, Nick Patterson, Morten Rasmussen, Robert Hastings, and Dienekes Pontikos for sharing their data with us and for fruitful discussions.  We also thank Alon Keinan and the Illumina development team for their feedback and support.

## Supplementary Materials

Supplementary Note 1 – attached in this document
Supplementary Note 2 – attached in this document
Supplementary Table 1 – is available from
http://eelhaik.aravindachakravartilab.org/files/TableS1.xlsx

## Materials and Methods

*Genotype Data Retrieval*

Genotyped samples for nearly 300 worldwide populations were obtained from 16 public and private collections [7,8,9,10,12,15,35,52,53,54,55,56,57,58,59,60] and the FamilyTreeDNA collection. To study gene flow from apes, ancient hominins, and modern humans, we used the dataset of 257,000 high-quality autosomal SNPs assembled by Reich et al. [61].

*Identifying AIMs*

Ancestral informative markers (AIMs) were collected using different approaches. To begin with, 55,000 AIMs were harvested from the published literature [2,16,62,63,64,65,66,67,68,69,70,71,72,73,74,75,76] and 12,000 AIMs were obtained from private collections., We also employed two methods, *infocalc* [30] and *AIMsFinder,* to find additional AIMs in our global multi-population panels. Because *infocalc* [77] does not estimate the minimal number of AIMs necessary to distinguish populations from each other, but rather ranks SNPs by their informativeness for inferring ancestry, we selected the top 1% of *infocalc* results when applied on our population panels.

To find the minimal set of AIMs necessary to distinguish any two populations, we developed *AIMsFinder*. *AIMsFinder* is a novel principal component (PC)-based approach that identifies the most informative set of markers that can distinguish between two populations. To circumvent biases caused by the comparison of uneven number of populations [78,79], we implemented Elhaik's [37] dual-population framework consisting of three "outgroup" populations that are available in large sample sizes and are the least admixed with each other - Mbuti and Biaka Pygmies (Africa), French Basques (Europe), and Han Chinese (Asia) - and two populations of interest, all of equal sample sizes. This framework minimizes the number of significant PCs to four or fewer (Tracy-Widom test, $p<0.01$) and maximizes the portion of explained variance to over 20% for the first two PCs. Convex hulls were calculated using Matlab "convhull" function and plotted around the cluster centroids. The relatedness between two populations of interest was estimated by the commensurate overlap of their clusters. The *AIMsFinder* attempts to find a minimal set of markers that reduces this overlap.

Principal components were calculated by first forming a matrix consisting of $m$ subjects (rows) and $n$ SNPs (columns) and sorting each pair of bases alphabetically. Next, the genotype data were transformed to an integer matrix $A$ of the same size, where each entry is encoded as 0, 1, 2, or empty based on the count of the left allele. In other words, if $B_1$ and $B_2$ were the bases to appear in the $j$-th SNP (in alphabetical order), $B_1$ homozygotes would be encoded as 0, $B_1B_2$ would be encoded as 1, and $B_2$ homozygotes would be set to 2. Missing SNPs were removed from all populations. Using the approach described by Paschou et al. [80], a singular value decomposition (SVD) was applied on *A* in order to compute its singular vectors and values. The SVD returns *m* nonnegative singular values *s*, *m* pairwise orthonormal eigenvectors *u*, and *n* pairwise orthonormal eigenvectors *v*. The SVD of the *m*x*n* matrix *A* is given by $suv^T$ and can be written as a sum of outer products:

$$A = \sum_{i=1}^{m} s^i u^i v^{iT}. \qquad (1)$$

The left singular vectors ($u^i$) are the linear combinations of the columns (SNPs) of the matrix *A* and are denoted *eigenSNPs*. In the dual-population framework, the number of significant principal components *k* corresponding to the *eigenSNPs* is less than four and for simplicity was set to two. The columns (SNPs) that correlate with the top *eigenSNPs* can be identified by rearranging Eq. 1 such that the *j*-th column (SNP) of matrix *A* (denoted by $A_j$) can be expressed as

$$A^j \approx \sum_{i=1}^{k} (s^i u^i) v_j^i, \qquad (2)$$

where $v_j^i$ is the *j*-th element of the *i*-th right singular vector and $A^j$ is the linear combination of the top *k* left singular vectors and corresponding singular values. Finally, SNPs were scored by $p_j$

$$p_j = \sum_{i=1}^{k} (v_j^i)^2 \qquad (3)$$

Paschou et al. [80] proposed to use all the inferred SNPs for further analysis, even though such a strategy would yield a very large number of SNPs. *AIMsFinder*'s strategy is to rank the SNPs by their score, iteratively select the top 50 SNPs, and test whether using only these SNPs will reduce the overlap between the two populations in question. If so, then the process is repeated with an additional 50 SNPs until either no further reduction is achieved or a threshold of 2,000 SNPs is reached. We carried out these pairwise calculations on all populations within the same continents for which genotype data were available. Less than 10% of these populations reached the threshold. Both *infocalc* and *AIMsFinder* were applied on the multi-population panels independently and collectively, and retrieved nearly 75,000 AIMs.

*Identifying regions of potential gene flow from archaic hominins into modern humans*

Several regions showing potential interbreeding between Neanderthals and Denisovan and modern humans were identified from existing datasets and the literature. First, we obtained ~160,000 SNPs from the USCS table *ntSssSNPs* [5], in which at least four of six modern human genomes (human reference, San, Yoruba, Han, Papuan, and French) have the derived allele while all observed Neanderthal alleles have the ancestral form. Because these SNPs were highly clustered (e.g., 89,000 SNPs are within 10,000 from one another), we pruned this dataset by filtering SNPs in high LD ($r^2$>0.8) and retaining only common SNPs (MAF >20%). Overall, we collected ~26,000 such SNPs. An additional ~600 SNPs were collected from similar regions identified by Noonan et al. [32]. We further included 200 SNPs within 13 candidate regions for gene flow from Neanderthal to non-African modern humans [5] in which both Neanderthal and Denisovan share the ancestral allele with non-African populations [61]. An additional 200 common (MAF >20%) SNPs embedded within and around these haplotypes, as well as 125 SNPs the X-chromosomal haplotype (B006) and its flanking regions [33], were also included.

New candidate regions enriched for alleles shared between Neanderthals and modern humans were detected in the following way. First, we identified all SNPs in which Neanderthals had the derived allele but chimpanzees and Denisovans carried the ancestral allele. Of these SNPs, we filtered out those in which Melanesians carried a lower frequency of the derived allele compared to all other populations. Next, we identified regions enriched for the Melanesian high frequency derived allele by modifying *IsoPlotter*, originally designed to identify compositionally homogeneous genomic domains [34]. In brief, if SNPs in the regions of interest are marked as 1's and all other SNPs are marked as 0's, then the algorithm finds clusters with high frequency of 1's in an unbiased manner [34]. We retained regions in which the derived alleles were of high frequency (>20%). The same analysis was carried out for the Melanesian low frequency derived allele SNPs. We repeated these analyses for alleles shared between modern humans, Denisovans, and chimpanzees. Overall, ~5,000 SNPs from candidate interbreeding regions were collected, giving us a total of ~30,000 SNPs dedicated to studying gene flow from archaic hominines to modern populations. In this regard, we emphasize that many SNPs are informative for more than one type of analyses so that the actual number of SNPs used for each analysis is higher.

*SNP validation*

To cross-validate the GenoChip's autosomal genotypes, we genotyped 168 samples from 14 worldwide populations of the 1000 Genomes Project including: Americans of African ancestry (Southwest USA), Americans of Mexican ancestry (Los Angeles, USA), Americans with Northern and Western European ancestry (Utah, USA), British (England and Scotland), Finnish (Finland), Gujarati Indians (Houston, USA), Han Chinese (Bejing, China), Iberian (Spain),

Italians (Tuscany, Italy), Japanese (Tokyo, Japan), Kinh (Ho Chi Minh City, Vietnam), Luhya (Webuye, Kenya), Peruvians (Lima, Peru), and Yoruba (Ibadan, Nigeria). We confirmed that at least one of the alleles matched those reported by the 1000 Genomes Project. The concordance rate between GenoChip and the 1000 Genomes Project genotypes was calculated as the proportion of genotypes that were identical between the two datasets.

*Comparing population genetic summary statistics between genotyping arrays*

To compare the performances of the validated 130,329 autosomal and X-chromosomal SNPs of the GenoChip array to commercial arrays, we obtained the list of SNPs for the Illumina Human660W-Quad BeadChip (544,366 SNPs) from Illumina and Affymetrix Axiom Human Origins array (627,719 SNPs) available at ftp://ftp.cephb.fr/hgdp_supp10/Harvard_HGDP-CEPH/all_snp.map.gz. Because of the lack of overlap between these genotyping arrays, we used subsets of data calculated for HapMap III populations. MAF and $F_{ST}$ estimates for African, European, and Asians, were obtained from the "continental" HapMap dataset, as described in Elhaik [49]. Briefly, genotype data of 602 unrelated individuals from eight populations (YRI, LWK, MKK, CEU, TSI, CHB, CHD, and JPT) were downloaded from the International HapMap Project web site (phase 3, second draft) [9], passed through rigorous filtering criteria, and finally merged into continental populations (African (288), European (144), and Asian (170)). The final continental data set consisted of SNPs genotyped in at least one population from each continent.

We followed Wright's [45] method to calculate $F_{ST}$. For each SNP, we calculated the frequencies of both ancestral and derived alleles in each population. We then identified the allele with the smallest global frequency ($P$) when calculated as a weighted average over all populations such that ($0 \le P \le 0.5$). The frequency of that allele was considered the minor allele frequency. Similarly, the variance of the minor allele frequency $\sigma_P^2$ was obtained and $F_{ST}$ was calculated as:

$$F_{ST} = \frac{\sigma_P^2}{P(1-P)} \quad (4)$$

The MAF and $F_{ST}$ values of the continental dataset for autosomal (2,823,367) and X-chromosomal (86,449) SNPs were compared to those obtained from GenoChip (126,425 and 2,421 SNPs), Illumina Human660W (541,104 and 12,916 SNPs, respectively), and Affymetrix Axiom Human Origins Array (308,949 and 2,984 SNPs, respectively).

## Figures

Figure 1. Ancestry informative markers from over 450 world populations were harvested from the literature (green) and from public and private collections (red) including over 30 Jewish populations (blue). Coordinates were obtained either directly from the studies [10,12], according to the approximate region reported by the authors, or by the country's capital city.

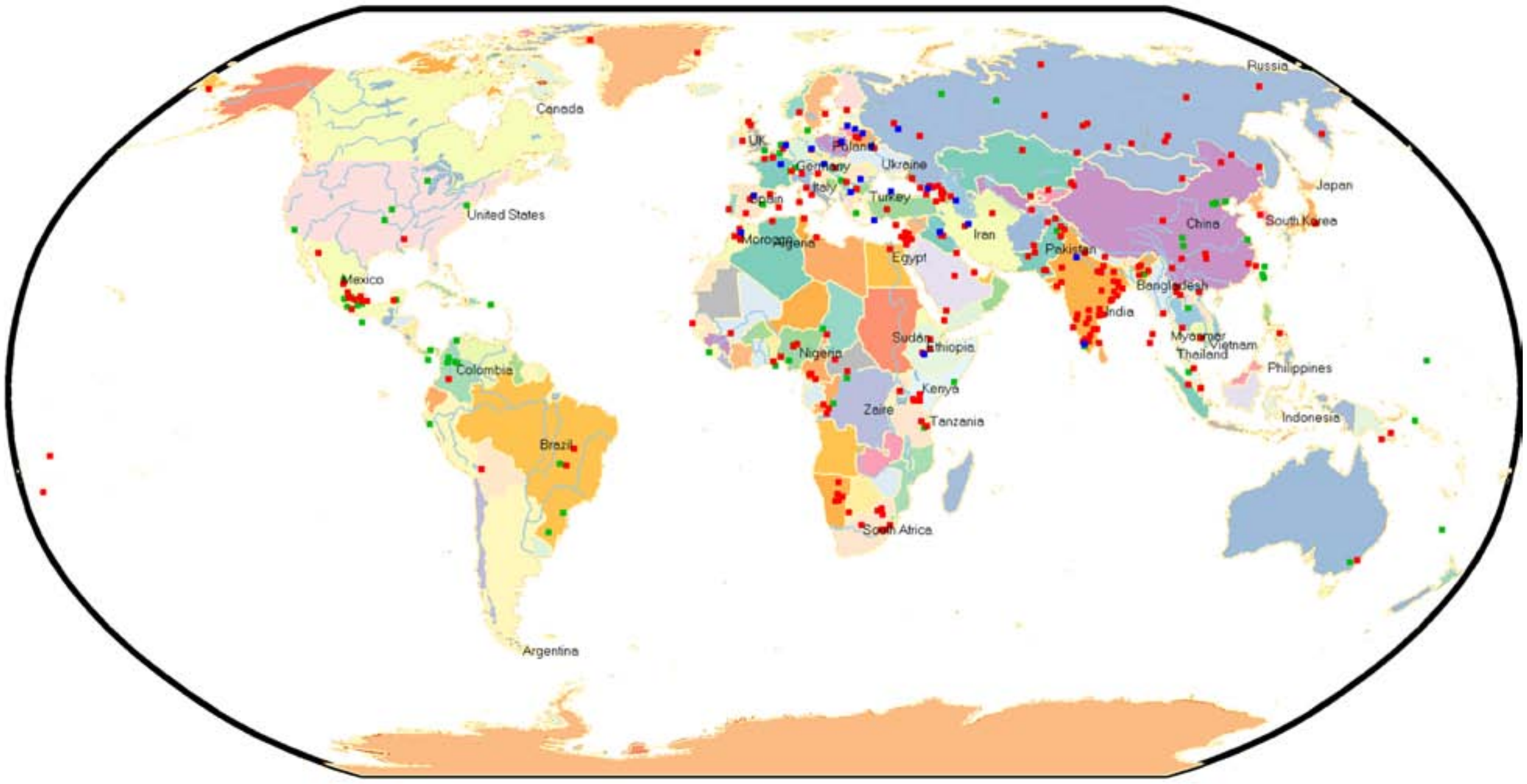

Figure 2.  SNP density in the Genochip.  The SNP densities across the genome are color coded to indicate the number of polymorphic SNPs per 100kb.  Gaps in the assembly are shown in gray.

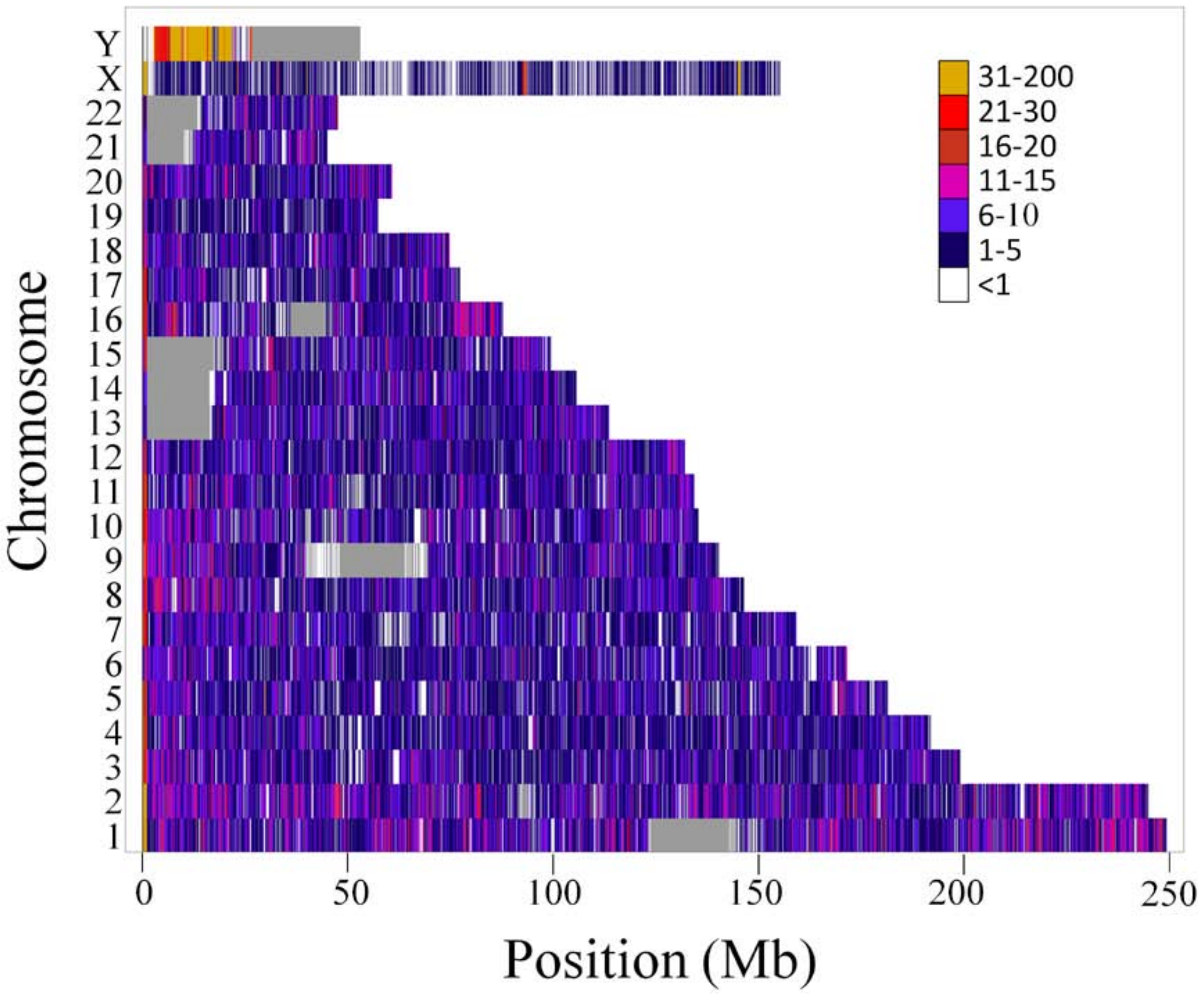

Figure 3. Success rate in validating SNPs defining Y-chromosomal and mtDNA haplogroups.

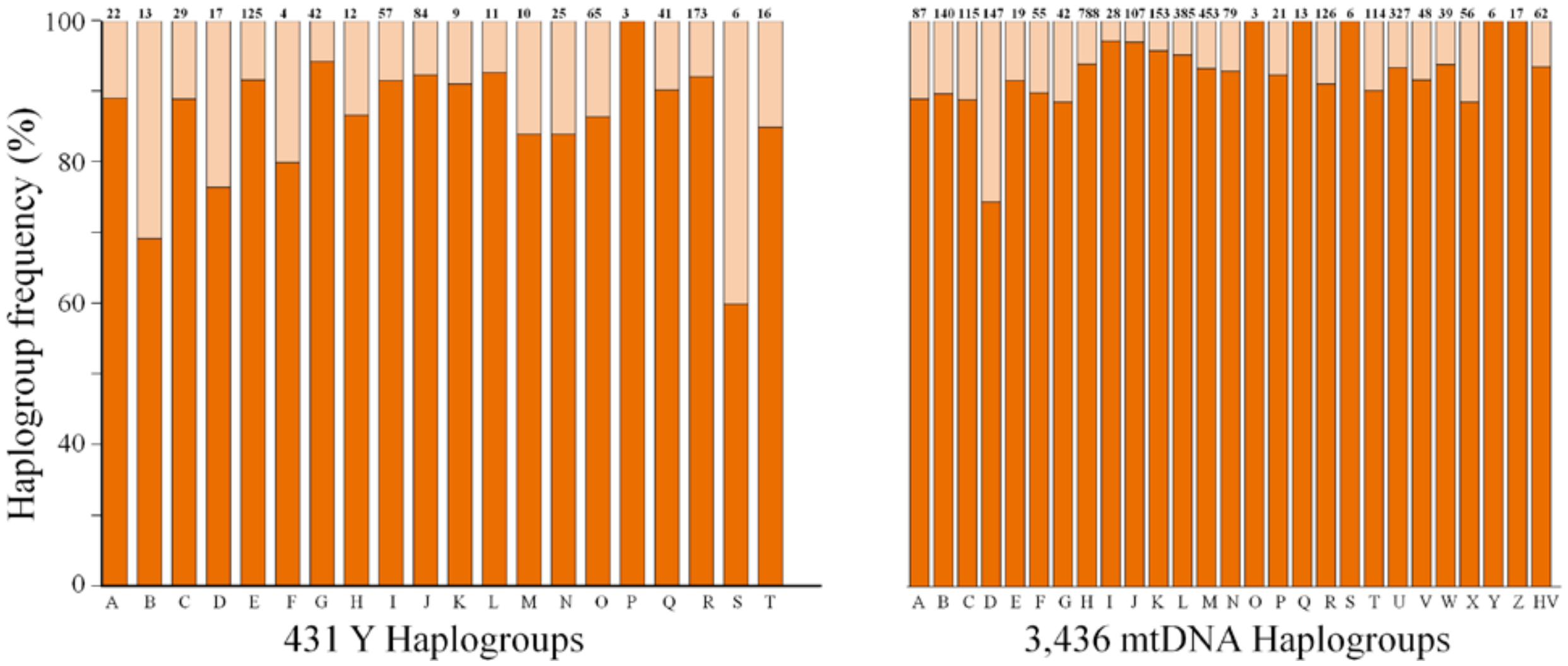

Figure 4. Minor allele frequency distributions for autosomal (a) and X-chromosomal (b) SNPs.

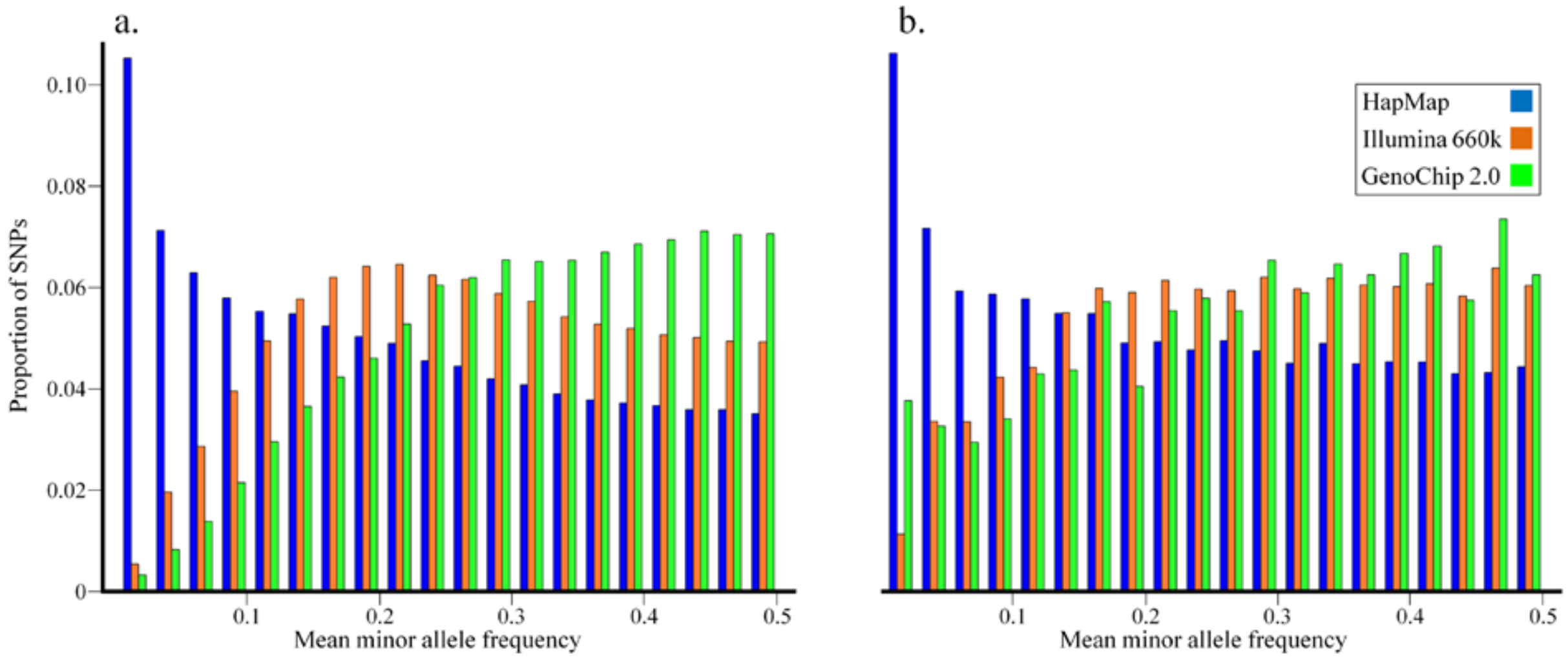

Figure 5. Distribution of locus-specific $F_{ST}$ in three continental populations. $F_{ST}$ values were obtained for (a) HapMap autosomal and (b) X-chromosomal SNPs. These values are compared to two subsets corresponding to Illumina Human660W and GenoChip SNPs. The histograms show bin distribution as indicated on the x-axis and the cumulative distribution (line).

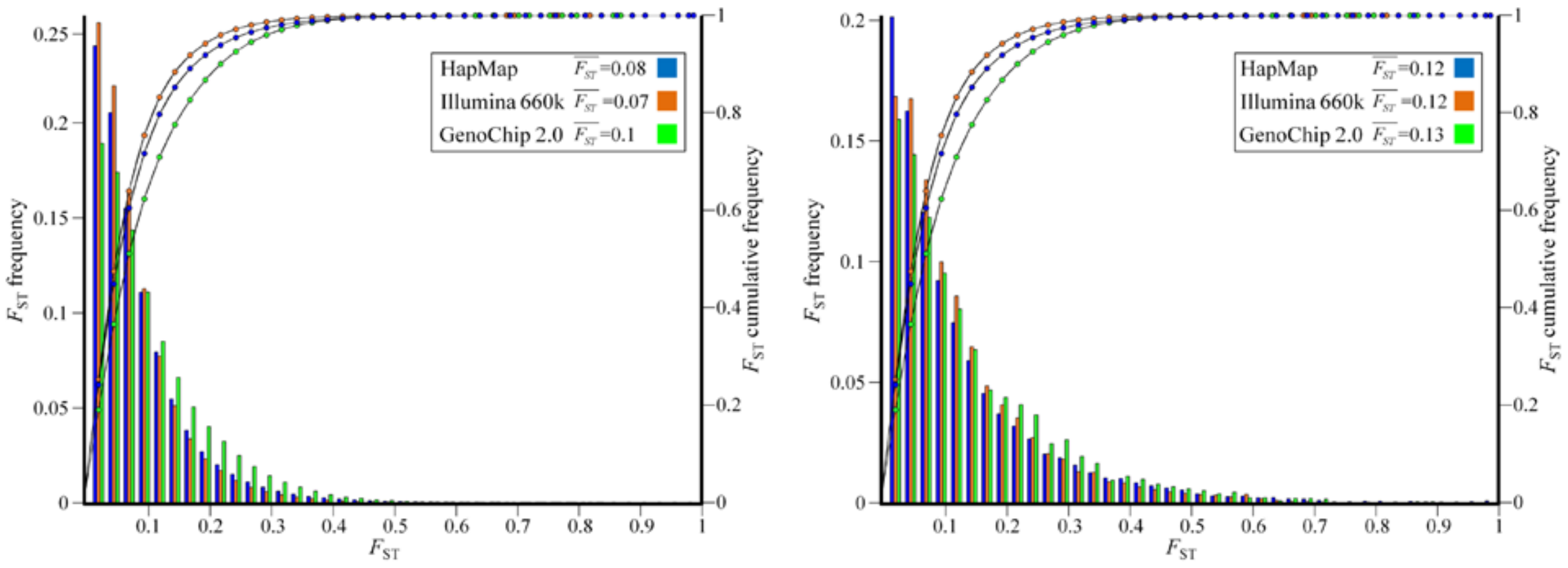

**Supplementary Figures**

Figure S1. Minor allele frequency distributions for autosomal (a) and X-chromosomal (b) SNPs.

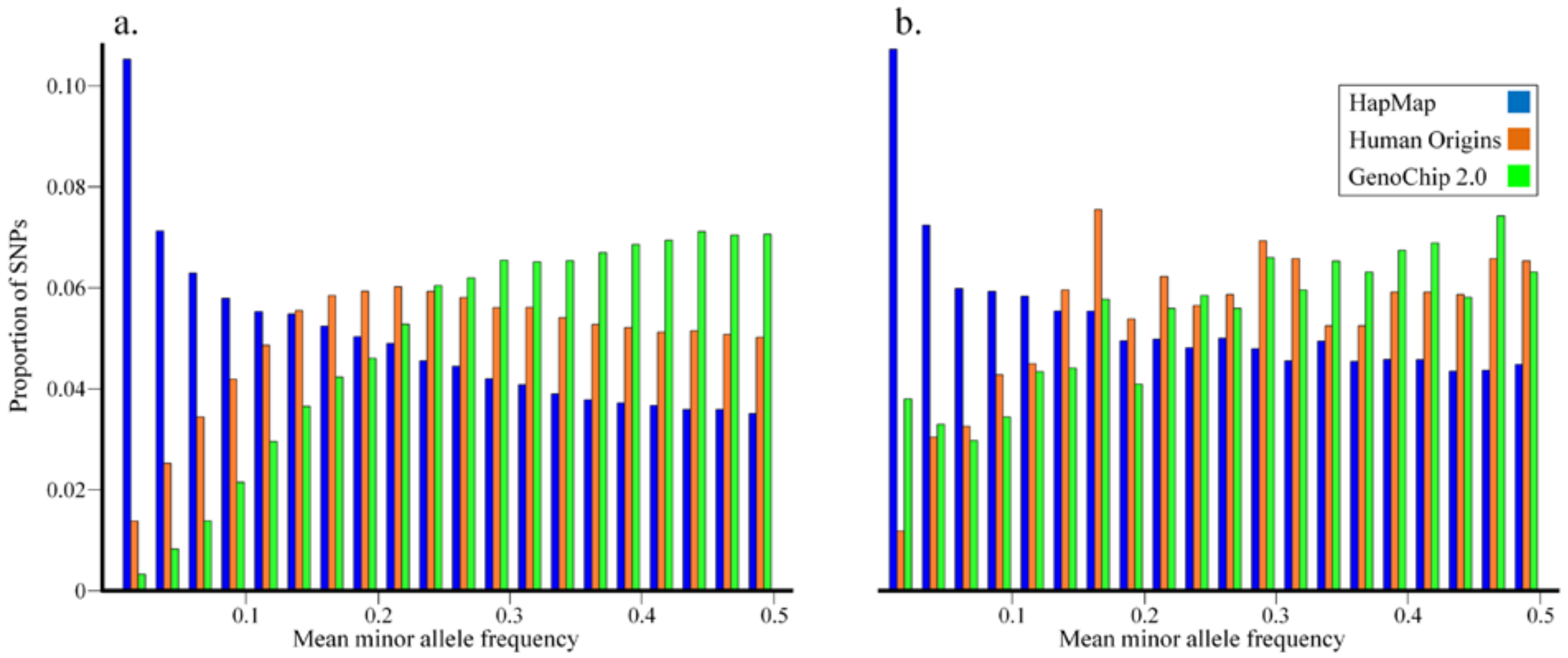

Figure S2. Distribution of locus-specific $F_{ST}$ estimates in three continental populations. $F_{ST}$ values were obtained for HapMap (a) autosomal and (b) X-chromosomal SNPs. These values are compared to two subsets corresponding to Human Origins and GenoChip SNPs. The histograms show bin distribution as indicated on the x-axis and the cumulative distribution (line).

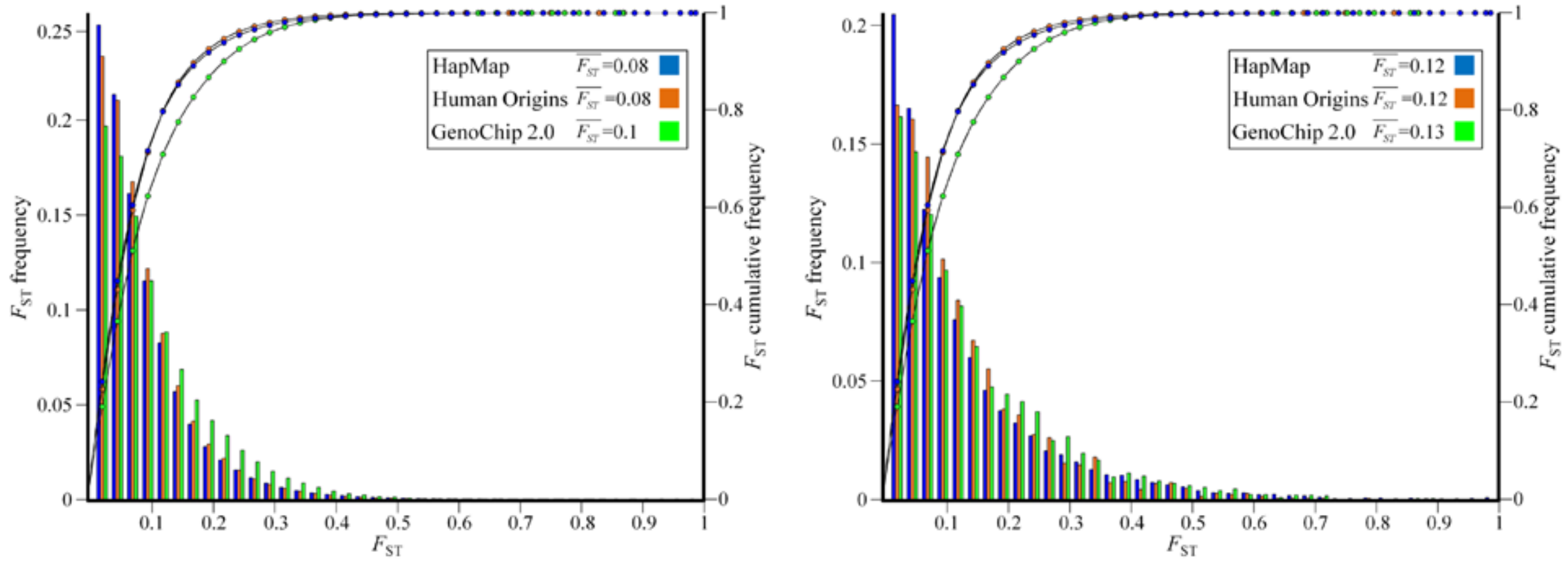

## Supplementary Text 1 - Y-chromosomal SNP inclusion

The Y chromosome SNPs were selected from multiple ongoing data collection efforts. The fundamental backbone of the Y chromosome tree comes from Karafet et al. [1], and the principal structure of this tree is still intact. The Y tree was expanded by including multiple published markers, most notably those from Rozen et al. [2] and Cruciani et al. [3]. Many of the SNPs that define additional branching points were found in the Family Tree DNA (FTDNA) laboratory by classical Sanger sequencing technology [4], where the new markers appeared on existing PCR amplicons that were used to type existing markers. Roughly 800,000 sequencing traces of Y chromosome specific PCR amplicons (250-700 bp each) have been screened for new markers.

FTDNA is maintaining a database of such inadvertently found mutations (http://ymap.ftdna.com) and keeps track of their phylogenetic position (http://ytree.ftdna.com). Moreover, most of FTDNA customers are actively interested in contributing the SNPs found in their genomic DNA to public research. For these reasons, we were able to add over 900 additional markers to the original data pool of 600 markers [1]. The phylogenetic positions of the novel markers on the Y tree have been verified and synchronized in close cooperation with the International Society of Genetic Genealogy (ISOGG, http://www.isogg.org).

In order to expand the Y tree resolution with more SNPs, FTDNA has undertaken a small number of Roche 454 sequencing runs. In short, the Y chromosome reference sequence (NCBI build 36) has been reformatted into a FASTA file of 1000 bp long subsequent segments. Those segments were Blasted against the whole human reference genome (NCBI build 36) to find matches to other chromosomes or other locations on the Y chromosome itself. The original Y chromosome segments (1000 bp) were sorted according to the expected value for the best non-identical Blast hit. The unique Y chromosome regions discovered in this way were submitted to NimbleGen [5] in order to design a custom specific 385k Sequence Capture Array, optimized for 454 Titanium sequencing. Genomic DNA from buccal swabs of three voluntary FTDNA associated males was extracted and a Titanium compatible library was created for each volunteer. Each library was enriched on the custom 385k Sequence Capture Array [6] and sequenced according to the Roche XL+ Sequencing Method Manual for the GS FLX+ Instrument [7]. Novel Y chromosomal SNPs were extracted from the reads with the help of the Roche GS Mapper software, verified by Sanger sequencing, and then added to the pool when confirmed.

In addition to this well studied set of 1985 core Y markers, we also included SNP candidate markers that have been found by mining public datasets, such as the 1000 Genomes Project [8] or second generation sequencing data from associated working groups that shared their data with us like PF, CTS, F (see Figure S1). After merging datasets from these multiple sources, we started from a raw SNP candidate database of approximately 27,500 markers. To consolidate these data sets, we first verified each dataset and eliminated duplicate SNPs. We also removed insertion and deletion markers wherever an alternative point mutation defined a synonymous branch on the Y tree. Second, to reduce the number of Y chromosome markers to the specified limit of approximately 15,000 SNPs, we sorted out the Y SNP candidates (but not the verified 1985 Y SNP core markers) with the lowest confidence scores using the Illumina chip design tool.

The final set submitted to generate the Illumina bead pools contained 15,733 Y chromosome SNPs. The sources of origin are visualized in Figure S1.

## Figures

**Figure S1**. Sources of the Y chromosomal SNPs that have been included on the Illumina Bead Array.  The total 15,733 Y markers resemble a merged collection of well-studied core markers (green) and a larger set of Y candidate SNPs (yellow).  The pale yellow color indicates that no phylogenetic information has been reported by the authors.

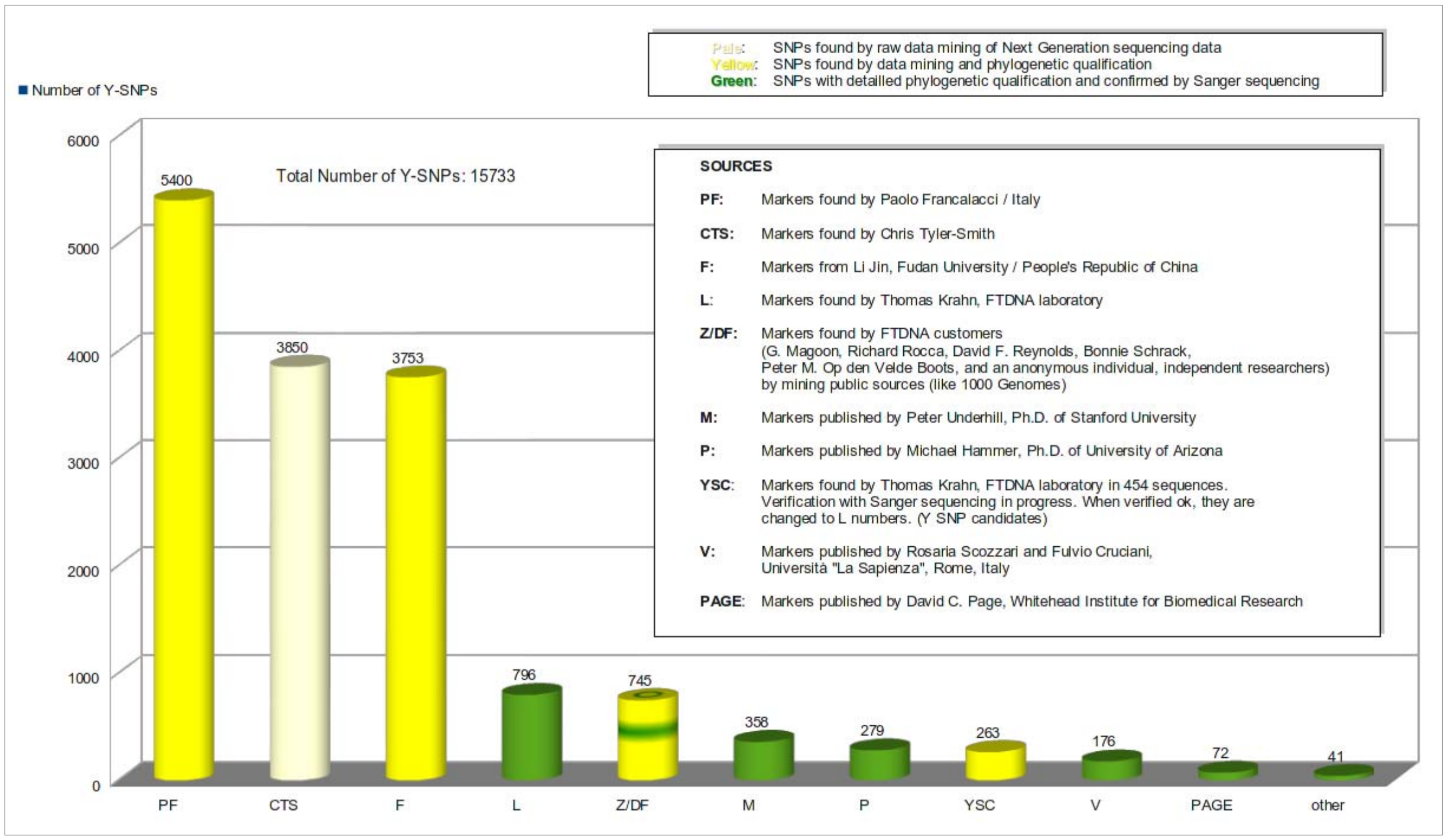

**Supplementary Text 2 – Mitochondrial DNA SNP inclusion**

To infer high-resolution mitochondrial DNA (mtDNA) haplogroups, we aimed to include highly informative mtDNA single nucleotide polymorphisms (SNPs) ascertained from a large number of samples. To achieve this goal, we combined the Genbank and the private FamilyTreeDNA databases to assemble a dataset of all 6,409 known mtDNA SNPs observed in 260,782 worldwide samples. These samples include all 1,277 known haplogroups, though this number was later redefined to be 3,552 (Behar et al. 2012).

Of these SNPs, 5,589 are located in coding regions and the remaining appear in the two main hypervariable regions (HVRs). All samples were genotyped for the HVR-1 region, 57,227 samples were also genotyped for the HVR-2 region, and 20,000 were genotyped for both HVR regions and the coding region.

The next step in the process was designing probes for the chip. The Illumina probe design process requires a minimum of 50 mutation-free (bp) immediately preceding or following the SNP of interest. Similar to the design of PCR primers, a probe has to be unique to the location harboring this SNP. While this process is relatively simple when working with autosomes, the mtDNA presents a thorny challenge due to its immense variability and mutation density. For example, of the total 1,118 positions in the control region, some 73% positions are mutable. This means that each position preceded by 50 SNPs would require $10^{15}$ probes to allow detection of these variants. Because capturing all mtDNA SNPs would require an imaginary number of probes, we developed an alternative approach to survey the mitochondrial genome for sequencing variation (Figure S2).

First, because of the low whole genome amplification hybridization temperature, high confidence levels of probe binding can be achieved with only 20 bases. Therefore, the number of mutation-free DNA bases adjacent to the SNP of interest was reduced from 50 to 20 bp. Second, for each SNP, we calculated the frequency of mutations in our dataset in order to exclude the most extremely rare SNPs. Accordingly, we included only SNPs with a frequency higher than $10^{-5}$ in the coding regions, $3 \cdot 10^{-4}$ in the HVR-2 region, and $7*10^{-5}$ in the HVR-1 region. Next, we developed a program to automate the probe design based on the Revised Cambridge Reference Sequence (rCRS) (Andrews et al. 1999). As probes can be designed from either side of a SNP, we increase our chances of success by having a different set of probes from each direction of each SNP, unless impossible due to the existence indels or poly-C regions. This layer of redundancy increased the success rate of the GenoChip and allowed us to identify SNPs and haplogroups commonly absent from commercial platform. For example, allele A in position 8,860 (haplogroup H2a2), residing in a highly mutable region, is a definite marker for Western Europeans. We were able to capture this SNP with nearly 600 probes. Another example involves haplogroup B, which, in being defined by a deletion event of 9-bp starting at position 8,281, is very difficult to capture. To overcome this problem, we designed a probe that identifies a transition event at the next base following the deleted region.

Overall, ~32,000 probes were designed for the GenoChip to identify ~3,800 SNPs that capture over 90% of all 3,552 known haplogroups (Behar et al. 2012) (Figure 2).

## Figures Legend

**Figure S2.** An illustration of the SNP inclusion process.

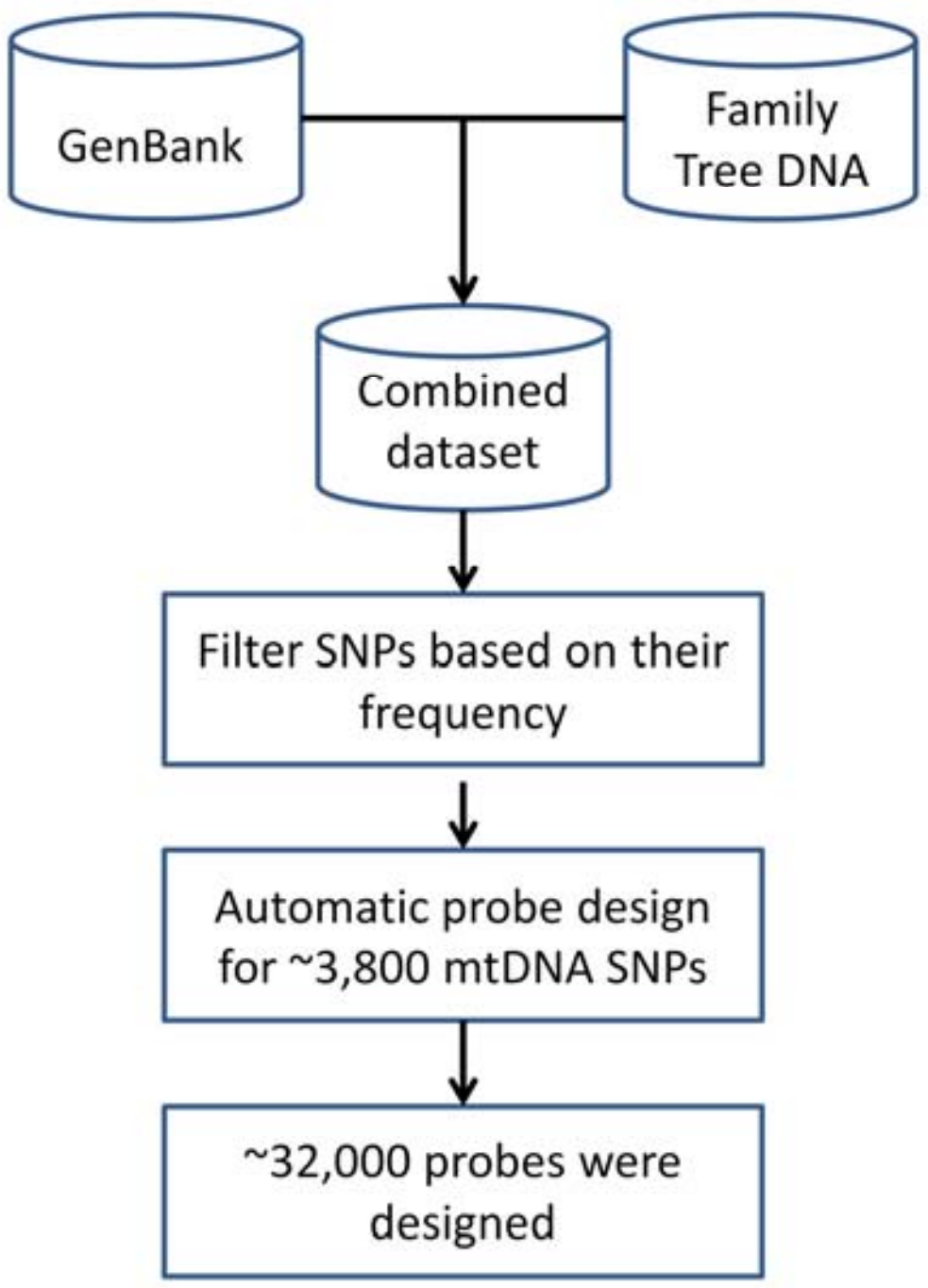